 \documentstyle[emulateapj,epsf,rotate]{article}

\def\bt{\begin{tabbing}}
\def\et{\end{tabbing}}
\def\beq#1{\begin{equation}\label{#1}}
\def\eeq{\end{equation}}
\def\eq#1{{\frenchspacing eq.}~(\ref{#1})}

\def\xbf{{\bf x}}

\def\ybf{{\bf y}}
\def\nbf{{\bf n}}

\def\Cbf{{\bf C}}
\def\alphaHat{\hat{\alpha}}
\def\da{\Delta\alphaHat}
\def\dar{\da_{\small{T}}}
\def\damc{\da_{\small{M}}}
\def\dT{\delta T}
\def\syn{synchrotron~}

\def\microk{$\mu$K~}
\def\mmicrok{$\mu$K}
\def\microm{$\mu$m~}

\def\mj{MJy/sr}

\def\rms{{rms}} 
\def\etal{{\frenchspacing\it et al.~}}
\def\eg{{\frenchspacing\it e.g.}}

\def\l{\ell}
\def\expec#1{\langle#1\rangle}

\def\spose#1{\hbox to 0pt{#1\hss}}
\def\simlt{\mathrel{\spose{\lower 3pt\hbox{$\mathchar"218$}}
     \raise 2.0pt\hbox{$\mathchar"13C$}}}
\def\simgt{\mathrel{\spose{\lower 3pt\hbox{$\mathchar"218$}}
     \raise 2.0pt\hbox{$\mathchar"13E$}}}
\def\simpropto{\mathrel{\spose{\lower 3pt\hbox{$\mathchar"218$}}
     \raise 2.0pt\hbox{$\propto$}}}

\def\pp{\noindent\parshape 2 0truecm 13.6truecm 1truecm 12.6truecm}
\def\rn{\pp}

\def\bfig{\begin{figure}[h] \centerline{\hbox{}}\vfill}
\def\efig{\end{figure}\vfill\newpage}

\def\fig#1{Figure~\ref{#1}}

 \journalid{337}{15 January 1989}
 \articleid{11}{14}

\lefthead{DE OLIVEIRA-COSTA ET AL.}
\righthead{GALACTIC EMISSION AT 19 GHZ}
 \slugcomment{Accepted for publication in ApJL}

\begin{document}

\title{GALACTIC EMISSION AT 19~GHz}

\author{Ang\'elica de Oliveira-Costa\altaffilmark{1,2,3}, 
	Max Tegmark\altaffilmark{3,4}, 
	Lyman A. Page\altaffilmark{1} \&
	Stephen P. Boughn\altaffilmark{5}
	}

\begin{abstract}
We cross-correlate a 19~GHz full sky Cosmic Microwave Background 
(CMB) survey with other maps to quantify the foreground 
contribution.  Correlations are detected with the 
Diffuse Infrared Background Experiment (DIRBE) 
240, 140 and 100\microm maps at high latitudes 
($|b|$$>$30$^{\circ}$), and marginal correlations are detected with the Haslam 
408~MHz and the Reich \& Reich 1420~MHz \syn maps.
The former agree well with extrapolations from higher frequencies 
probed by the COBE DMR and Saskatoon 
experiments and are consistent with both free-free and rotating dust 
grain emission.
\end{abstract}

\keywords{cosmic microwave background --  methods: data analysis}


\section{INTRODUCTION}
 
One of the major challenges in any Cosmic Microwave Background (CMB) 
anisotropy analysis is to determine the fraction of the observed signal due
to diffuse Galactic emission.  Three components of Galactic emission have been
firmly identified: \syn and free-free radiation, which are important mainly at 
frequencies below 60~GHz, and thermal emission from dust particles, which is 
important mainly at frequencies above 60~GHz (see, \eg, Weiss 1980;  Bennett 
\etal 1992; Brandt \etal 1994; Tegmark and Efstathiou 1996). 
In principle, these three components can be discriminated
by their frequency dependence and morphology.
In practice, however, there is no component for which both the 
frequency dependence and the spatial template are currently well known 
(see, \eg, Kogut \etal 1996a, hereafter K96a, and references therein). 

The cross-correlation technique allows one to estimate the {\rms} level of
Galactic emission present in a CMB map.
For instance, an analysis of high latitude Galactic emission in the 
COBE DMR map gave {\rms} estimates of 
(3.4$\pm$3.7)~\microk for \syn and (2.7$\pm$1.3)~\microk for dust emission 
at 53~GHz on a 7$^{\circ}$ scale (Kogut \etal 1996b, hereafter K96b). 
A third component, correlated with the DIRBE maps but decreasing with 
frequency, was detected at the level of (7.1$\pm$1.7)~\microk and 
tentatively identified as free-free emission (K96b).
Although this component was also detected in the Saskatoon maps 
(de Oliveira-Costa \etal 1997, hereafter dOC97) on a 1$^{\circ}$ scale, 
at a level (17.5$\pm$9.5)~\microk at 40~GHz, it is still not clear 
if it is due to free-free emission.
An analysis of Owens Valley Radio Observatory (OVRO) data at 14.5 GHz 
showed such a component at the level of 203~\microk in a small sky region
on scales of 7'-22' (Leitch \etal 1997, hereafter L97), but this level 
is substantially higher than free-free emission estimates based on 
H$\alpha$ images (Gaustad \etal 1996; Simonetti \etal 1996).
Suggested explanations include the presence of high temperature gas 
(L97) and rotating dust
grains (Draine \& Lazarian 1998, hereafter DL98), 
but more data 

\bigskip
        {\footnotesize
$^1$Princeton University, Dept. of Physics, Princeton, NJ 08544 

$^2$angelica@ias.edu 

$^3$Institute for Advanced Study, Olden Lane, Princeton, NJ 08540 

$^4$Hubble Fellow

$^5$Haverford College, Dept. of Astronomy, Haverford, PA 19041
        }
\goodbreak

   \medskip
   \centerline{\rotate[r]{\vbox{\epsfxsize=12.20cm\epsfbox{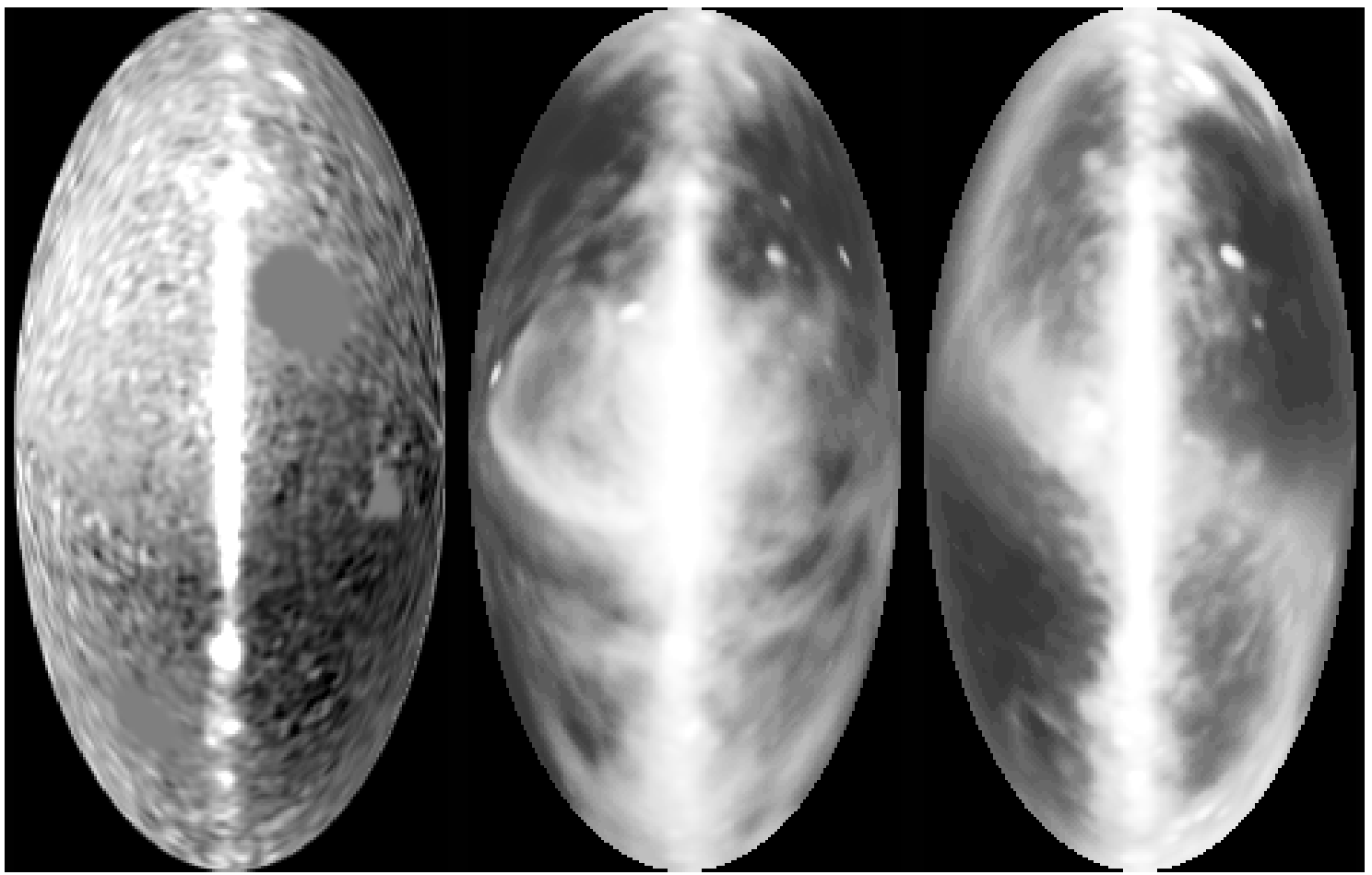}}}}
   \figcaption{19 GHz survey (top), 408 MHz Haslam
   synchrotron template (middle), and DIRBE 100\microm dust template
   (bottom).
   \label{MapFig}
   }

\bigskip
\noindent
is needed to settle this issue. 
The purpose of this letter is to evaluate
the Galactic contribution to the 19~GHz full sky map 
(see \fig{MapFig}) 
by cross-correlating it
with the DIRBE dust maps, and with the Haslam and Reich \&
Reich synchrotron maps. 

\section{METHOD}

The 19~GHz map consists of $N=24576$ pixels with sky temperatures
$y_i$ and noise $n_i$.  We assume that this map is a superposition of CMB
fluctuations and a Galactic component whose angular distribution 
is traced in part by an external data set.  Representing these 
contributions by $N$-dimensional vectors,
\beq{signals}
  	\ybf = \nbf + \xbf_{CMB} + {\alpha} \xbf_{Gal} + \ybf_{Gal},
\eeq
where $x_{CMB}^i$ is the contribution of the fluctuating component
of the CMB, $x_{Gal}^i$ is the brightness fluctuations 
of the Galactic template map (not necessarily in temperature units), 
$\alpha$ is the coefficient that converts units of the Galactic
template into antenna temperature and $y_{Gal}^i$ represents any 
residual Galactic contribution which is uncorrelated with $x_{Gal}^i$. 
We consider $\nbf$ and $\xbf_{CMB}$ to be random variables 
with zero mean, i.e.
	$\langle \xbf_{CMB} \rangle  = \langle \nbf \rangle  = 0$, 
	and 
	$\xbf_{Gal}$ and $\ybf_{Gal}$ 
to be constant vectors.  Thus the data covariance matrix is given by
\beq{varCMB}
  \Cbf \equiv 
       \langle  \ybf \ybf^T \rangle   - 
       \langle  \ybf \rangle \langle \ybf^T \rangle =
       \langle  \xbf_{CMB} \xbf_{CMB}^T \rangle + 
       \langle  \nbf \nbf^T \rangle,
\eeq
where $\langle \xbf_{CMB} \xbf_{CMB}^T \rangle $ is the covariance 
matrix of the CMB and $\langle \nbf \nbf^T \rangle $ is the noise 
covariance matrix. 
The noise in the 19~GHz map is approximately uncorrelated and has an
{\rms} amplitude of $\sigma_i$$\sim$2~mK.
Therefore, the covariance matrix of this map is
\beq{varCMBfinal}
 C_{ij}\approx\expec{n_i n_j}\approx\sigma_i^2 \delta_{ij}.
\eeq

Minimizing
$ \chi^2 \equiv 
          (\ybf - {\alpha} \xbf_{Gal})^T 
          {\bf C}^{-1}
          (\ybf - {\alpha} \xbf_{Gal}) $
yields the minimum-variance estimate of $\alpha$, i.e.
\beq{alpha}
   \alphaHat  = 
   \frac{\xbf_{Gal}^T ~ {\bf C}^{-1} ~  \ybf}
	{\xbf_{Gal}^T ~ {\bf C}^{-1} ~  \xbf_{Gal}}
\eeq
with variance 
\beq{varalpha}
   \da^2  = 
   \frac{1}{(\xbf_{Gal}^T ~ {\bf C}^{-1} ~ \xbf_{Gal})}.
\eeq
Note that unlike the case in dOC97, there is little 
contribution from chance alignments between the CMB and the various 
template maps, since the CMB contribution to ${\bf C}$ is negligible.
If the the noise is correlated or the anisotropy signal is
significant, then our $\Cbf$ used above will differ from the
true covariance matrix, denoted $\Cbf'$.
Although \eq{alpha} still provides a reasonable and unbiased estimate of 
	$\alpha$\footnote{
	Note that the diagonal approximation of 
	\eq{varCMBfinal} reduces \eq{alpha} to 
	a simple noise-weighted least-squares fit.
	},  
its variance will be larger than implied by \eq{varalpha}, given by
\beq{newvaralpha}
   \da^2  = 
   \frac{(\xbf_{Gal}^T ~ {\bf C}^{-1} ~ {\bf C}' ~ {\bf C}^{-1}~ 
   \xbf_{Gal})}{(\xbf_{Gal}^T ~ {\bf C}^{-1} ~ \xbf_{Gal})^2}.
\eeq
In the next section, an estimate of 
${\bf C'}$ is made from the data and the corresponding variance 
of ${\alphaHat}$ is evaluated.

\section{DATA ANALYSIS AND RESULTS}

The 19~GHz map has an angular resolution of 3$^{\circ}$ FWHM 
and is stored in $1.3^\circ \times 1.3^\circ$ pixels (Cottingham 1987; 
Boughn \etal 1992).  
The template maps are convolved with a 3$^{\circ}$ Gaussian beam and 
regions within 20$^{\circ}$ and 30$^{\circ}$ of the Galactic plane are 
excluded.  To avoid contamination by zodiacal dust emission, data within 
10$^{\circ}$ of the Ecliptic plane are also excluded from the analysis;
although, the results are found to be independent of this cut.
Off the Galactic plane, the 19~GHz map is dominated by the CMB dipole ($\l$=1).
On the other hand, because of its planar structure, emission
associated with the Galaxy has a strong quadrupole ($\l$=2) component.
Therefore, the monopole, dipole, and
quadrupole moments are removed from both the 19~GHz and template 
maps.
As a consequence, the computed $\alphaHat$ depends only on correlated structure
in the maps on angular scales $\simlt 90^{\circ}$.
  
\subsection{\it Correlations and their Error Bars}

The 19~GHz map was cross-correlated with five different templates: two for
\syn emission, the 408~MHz (Haslam \etal 1981) and 1420~MHz (Reich and Reich 
1988) surveys; and three to study dust and free-free emission, the
100, 140 and 240\microm Diffuse Infrared Background Experiment (DIRBE) sky 
maps (Boggess \etal 1992).  Table~1 lists the coefficients 
$\alphaHat$ derived from \eq{alpha} with errors computed from
\eq{varalpha}. 
All three DIRBE templates show significant correlations with the 19~GHz map, 
while the two \syn templates are found to be only marginally correlated.

Also listed in Table~1 are
the implied fluctuations in antenna temperature in the 19~GHz map, i.e.
$\dT = \alphaHat \sigma_{Gal}$, where $\sigma_{Gal}$ is the 
{\rms} of the template map.  If a template map includes a 
distinct component that is uncorrelated with the 19~GHz map, then  
$\dT$ is underestimated by a factor
$\sigma^{\prime}_{Gal} / \sigma_{Gal}$, where $\sigma^{\prime}_{Gal}$ is the 
{\rms} of the correlated component of the template map.
For this reason, the $\dT$'s in Table~1 should be considered lower limits.

As mentioned above, the errors listed in Table~1 were computed from 
\eq{varalpha} and are therefore lower limits to the error. We now 
describe a series of tests, performed to estimate the uncertainty 
in $\alphaHat$ due to correlated noise and other systematics.  

\subsubsection{\it The Noise Correlation Function}

If the true noise correlation matrix is isotropic,
	$C'_{ij} = \expec{n_in_j} = 
	           R(\theta_{ij}) \sigma_i \sigma_j$
where $\theta_{ij}$ is the angle between pixels $i$ and $j$,
then we can estimate the noise correlation function
$R(\theta)$ of the 19~GHz data (after removing the monopole, 
dipole, and quadrupole) by
	$R(\theta) = N_\theta^{-1} 
	             \sum_{ij} y_i y_j /\sigma_i \sigma_j$,
where $y_i$ is the antenna temperature of the $i^{th}$ 
pixel and the sum is over all $N_\theta$ pairs of pixels separated 
by $\theta$.  Substituting these relations into \eq{newvaralpha} 
gives estimates of $\da$ which are from 6\% to 17\% larger than 
those in Table~1. 

\subsubsection{\it Monte Carlo Simulations}

As another test of the robustness of the estimates of $\alphaHat$,
we employed Monte Carlo simulations in which the template maps are 
sliced into eighteen regions of equal area, each corresponding to a 
range of Galactic latitude $|b|$. Inside each of these regions the 
pixels are rearranged in random order, so that the latitude 
dependence is preserved but the longitudinal correlations are 
destroyed. Repeating this procedure 1000 times yields distributions 
of $\alphaHat$'s consistent with zero mean and with standard 
deviations $\damc\sim 1.6$ times larger than the formal errors 
$\da$ (see Table~1). 
	
\subsubsection{\it Sky Rotations}

Because of the approximate axial symmetry of the Galaxy, it is natural
to ask if the correlations are simply due to overall large-scale Galactic 
structure 
common to all emission components. To test this 
hypothesis, we repeated the analysis with 2 $\times$ 2 $\times$ 36 = 
144 transformed maps, rotated around the Galactic axis by multiples of 
10$^{\circ}$ and/or flipped vertically and/or horizontally.

For a 20$^{\circ}$ Galactic cut, the correct template map has the highest 
of all 144 correlations, while the distribution of these correlations has
standard deviation $\dar\sim 2.5$ times larger than the formal errors $\da$.
Likewise, the correct DIRBE maps have the highest of all 144 correlations
and the standard deviations are $\dar\sim \damc$, even for a 30$^{\circ}$ 
Galactic cut (see Table~1).
In contrast, we find no significant correlation between the 100\microm 
and Haslam maps, indicating that synchrotron and dust emission are not 
strongly correlated at high latitudes.

Since both $\damc$ and $\dar$ have 
correlated signal contributing to the noise, they constitute
overestimates of the true error bars.

%

\subsection{Latitude Dependence}

To investigate the dependence of the correlation on Galactic latitude,
we sliced the maps into six regions of equal area, each corresponding
to a range of latitude $|b|$, \fig{SliceFig} shows the results for the
100\microm map.
Note that $\alphaHat$ from the 19~GHz DIRBE correlation is almost 
the same at each latitude band, indicating that this correlation  
is not dominated by one or two nearby clouds.
The 19~GHz Haslam correlation is found to be more concentrated in 
the Galactic 
	plane\footnote{
	%
	In order to test how important spatially localized features 
	are for the correlation, we cross-correlated 19 GHz with 
	Haslam using only one hemisphere.
	Most of the correlation was found to come from 
	the northern hemisphere, presumably 
	because of emission features such as the North Polar Spur.
	}.

\vskip-3.1cm
\centerline{{\vbox{\epsfxsize=9.7cm\epsfbox{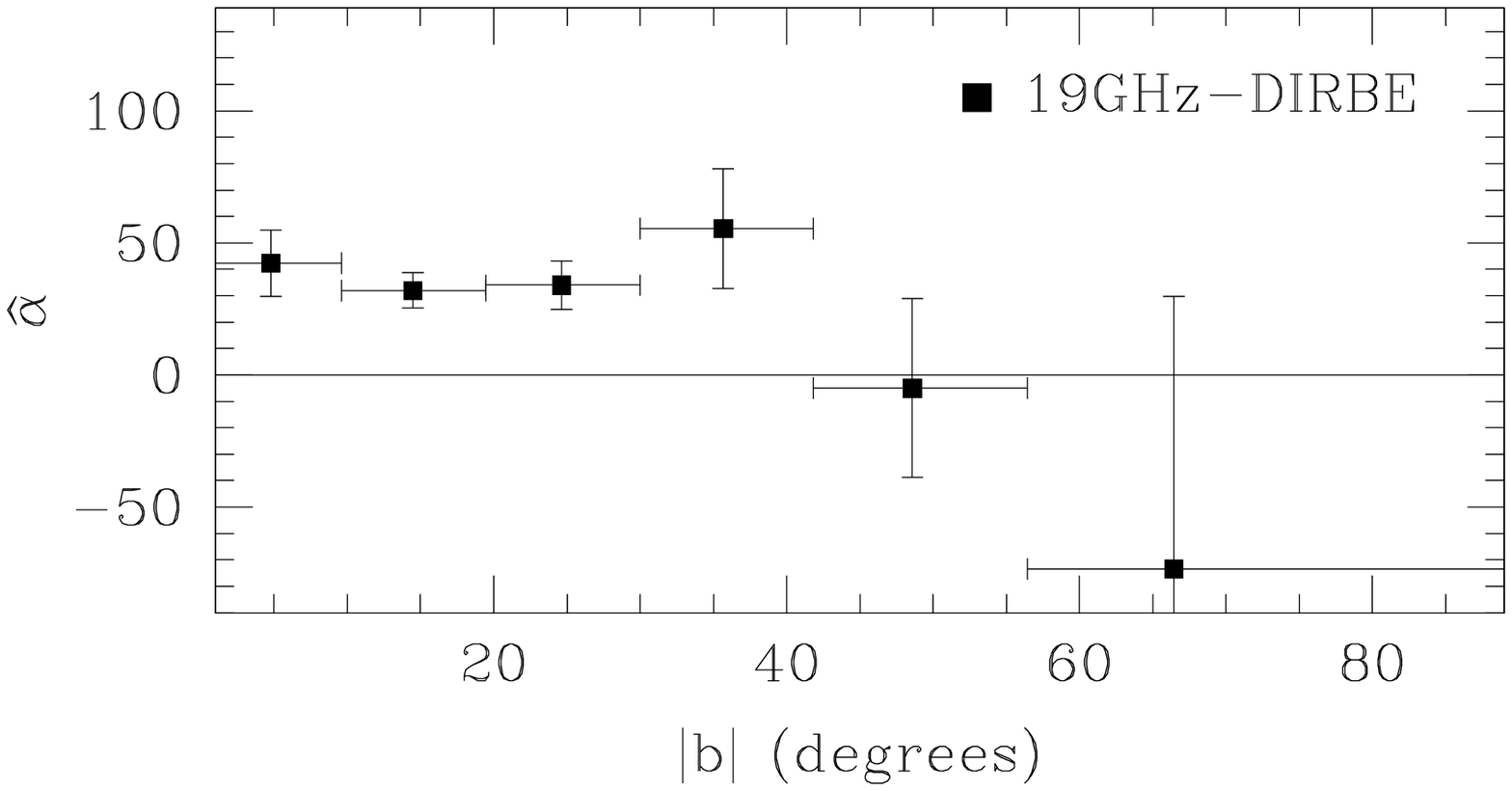}}}}
\vskip-1.7cm
	\figcaption{
	The dependence of $\alphaHat$ on Galactic latitude 
	$|b|$ for the 100\microm map. 
	The error bars are given by $\dar$.  
	\label{SliceFig}
	}
\smallskip
\smallskip

\subsection{\it Scale Dependence}

One way to determine angular scale of the correlation
is to compute the angular cross-correlation function (CCF) 
of the 19~GHz map with the template maps.  We computed this 
according to
  $CCF(\theta) = \left[\sum_{ij} x_{Gal,j} y_i / \sigma_i^2\right]/
                 \left[{\sum_{ij}\sigma_i^{-2}}\right]$,
where the sums are again over all pairs of pixels separated by 
$\theta$.

As an example, \fig{ccfFig} shows the $CCF(\theta)$ of 19~GHz and 
140\microm map, as well as the auto-correlation function of 
140\microm map.
It is clear that the correlated structure in the two maps is on 
small angular scales ($\theta\simlt 10^\circ$) and that the CCF 
is well-behaved on all angular scales.  
The latter gives us additional confirmation that there are no
large, unknown systematics that compromise the analysis.  Similar 
results are found for the other two DIRBE maps. The lower 
signal-to-noise for the correlation of the two radio templates 
result in CCF's which are less demonstrative. Note
that $CCF(0) \propto \alphaHat$.

\vskip-2.3cm
\centerline{\vbox{\epsfxsize=9.7cm\epsfbox{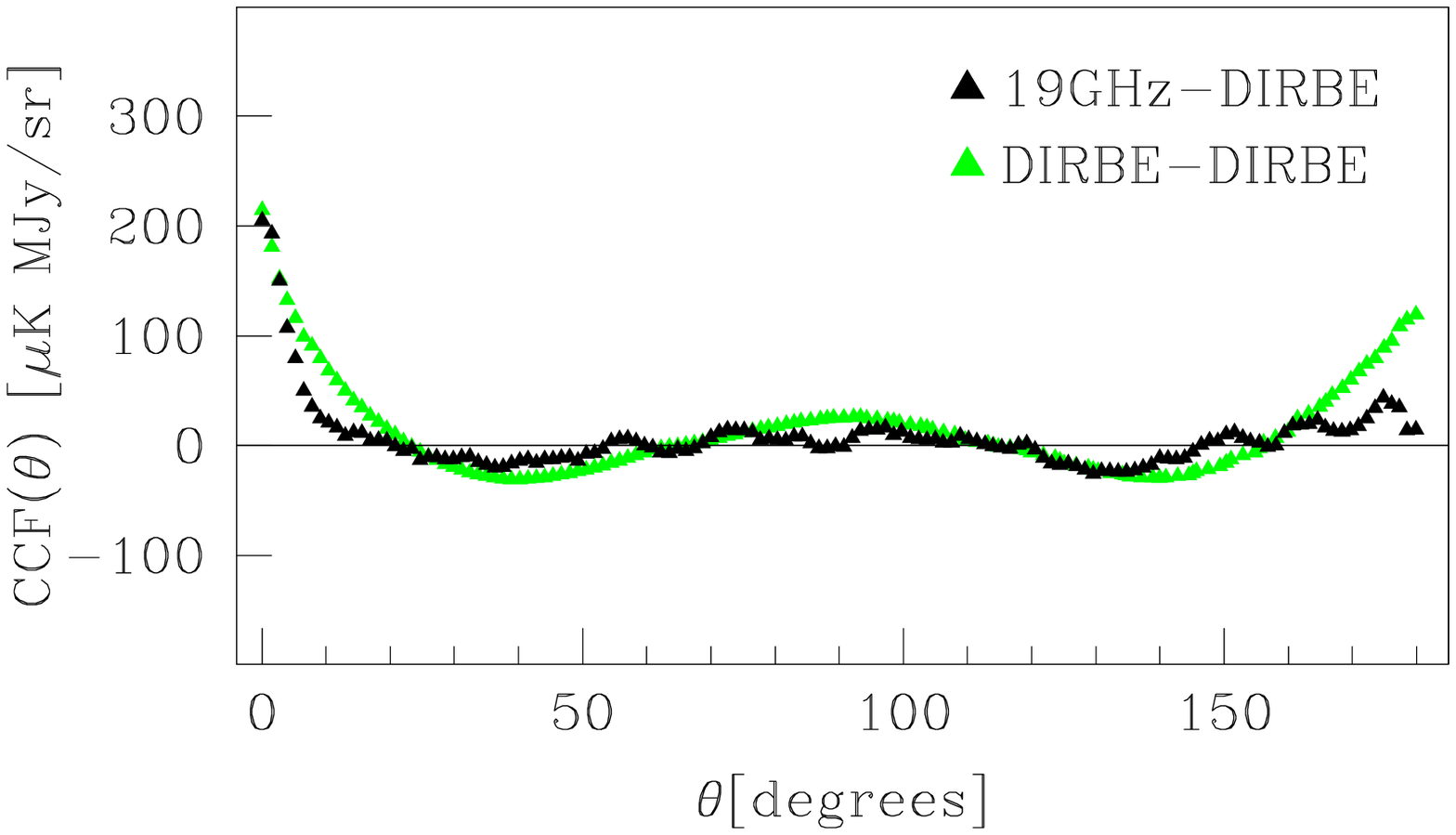}}}
\vskip-2.0cm
	\figcaption{
	CCF($\theta$) between 19~GHz and 140 \microm map 
        for a $30^\circ$ Galactic cut, together with the
	140 \microm auto-correlation.
	\label{ccfFig}
	}
\smallskip
\smallskip
\smallskip

As another way to investigate the dependence of the correlation on the 
angular scale, we repeated the analysis after high-pass filtering the 
19~GHz map using a partial sky multipole removal technique (see Tegmark 
\& Bunn 1995). Only after removing spherical harmonic components with 
$\ell\simgt 8$ does the correlation significantly decrease, which again 
indicates that large-scale Galactic structure is not the source of the 
correlation.
Both types of filtering therefore indicate that the bulk of the 
correlations are caused by fluctuations on scales of several degrees,
which is consistent with the shape of the CCF in \fig{ccfFig}.


\section{CONCLUSIONS}

The two \syn templates are found to be marginally correlated with
the 19~GHz map, while all three DIRBE far-infrared templates show a 
significant correlation.  When a \syn and a
DIRBE template are simultaneously fit to the 19~GHz map (via linear regression)
the correlation coefficients do not change significantly from those listed in
Table 1 and, in addition, the two fit parameters are essentially uncorrelated 
($|\rho|\simlt 0.05)$.  We conclude that the correlations with DIRBE dust
emission are independent of the correlations with \syn emission.  
The amplitude of the signal is much larger than expected 
for ordinary (vibrational) dust emission, as shown in \fig{ForegFig}.
Moreover, there is now good agreement between different experiments that
this correlated component is brighter at lower frequencies.

So what physical component is this? Two contenders have been proposed.
K96ab argue on physical grounds that free-free emission might be 
spatially correlated with dust. However, the correlations between 
H$\alpha$ (which is normally a good tracer of free-free emission) and CMB 
maps, and between~ H$\alpha$~ and the DIRBE maps, are weak~ (L97; 

\centerline{\hbox{$\>$}}
\vskip-1.8cm
\centerline{\vbox{\epsfxsize=9cm\epsfbox{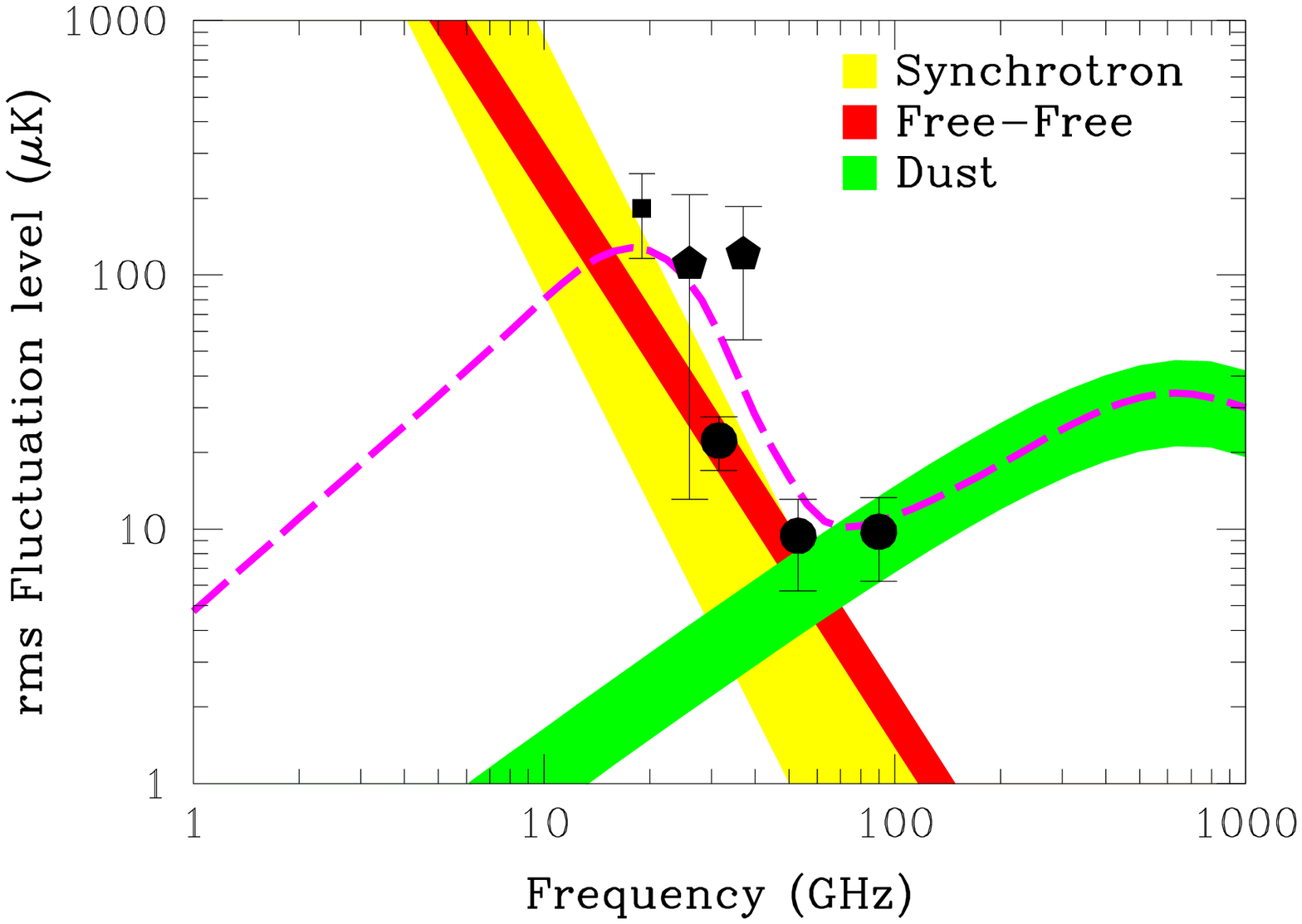}}}
\vskip-1.2cm
	\figcaption{
	Frequency dependence of the fluctuating component of 
	the Galactic emission for $|b|>30^\circ$. 
	The DMR correlations are represented by filled circles (K96b), 
	Saskatoon data by pentagons (dOC97) and 19~GHz by a square 
	with error bar given by $\dar$ (this work). 
	Dust emission (vibrational modes only) is normalized 
	using the DMR correlation at 90 GHz (K96a).
	The free-free model is normalized to the DMR correlation 
	at 31.5 GHz (the {\rms} free-free emission derived from 
	H$\alpha$ is much smaller than this DIRBE-correlated component).
	The \syn emission is normalized to the correlation between 
	the 19~GHz map and the 408~MHz Haslam map (see Table~1), 
	and is consistent with the 31.5 GHz upper limits from K96b.
	The thickness of each curve corresponds to the normalization 
	uncertainty. The dashed line is an example of emission 
	from rotating dust grains (DL98). 
	All fluctuations have been converted to DMR angular scales
	as $\dT = \varepsilon \alphaHat\sigma_{Gal}^i$, 
	where $\sigma_{Gal}^i$ is the rms of each template map and
	the correction factor is\\
	\centerline{
	  $\varepsilon = \left(
	  		  \left[\sum (2\ell+1)\ell^\beta W_\ell^{DMR}\right]/
	                  \left[\sum (2\ell+1)\ell^\beta W_\ell^i\right]
			  \right)^{1/2}
			  $.}\
        Here $W_\l^{DMR}$ is DMR window function, $W_\ell^i$
	is the window function of the experiment to be converted
	and we assume a $\beta=-3$ power spectrum slope. 
	\label{ForegFig}
	}
\bigskip

\noindent 
Kogut 1997; McCullough 1997).  One possibility is the presence of an extremely
hot ($\simgt$10$^6$K), ionized plasma (L97); however, DL98 have argued that
this cannot be true across the whole sky on energetic grounds.
These results motivated DL98 to suggest that perhaps it is dust 
after all, but emitting through rotational rather than vibrational 
excitations. As shown in \fig{ForegFig}, this can give a spectrum quite similar 
to that of free-free emission in the relevant frequency range.

H$\alpha$ maps with better accuracy are currently being made by 
the Dennison, Gaustad and Reynolds groups and should help to settle the issue. 
Another test is indicated by \fig{ForegFig}: although both the
rotating dust and the free-free models appear consistent with the
available 
	data\footnote{
        We compute the spectral index between 19 and DMR 53 GHz to be 
        $\sim -2.4$, and between 19 GHz and 1420 MHz to be 
	$\sim -2.8$.
	}, 
the former predicts a bump and a downturn around 10-20 GHz while the 
latter predicts a continued rise toward lower frequencies. 
A cross-correlation analysis with lower frequency data like 
the 10 GHz Tenerife map 
may be able to discriminate 
between these two models.

\bigskip

We would like to thank 
Ed Cheng,
Dave Cottingham,
Bruce Draine,
Dale Fixsen, 
Ken Ganga,
Ed Groth,
Al Kogut,
Alex Lazarian,
George Smoot and  
David Wilkinson for helpful comments.
Support for this work was provided by NASA grants NAG5-6034 and NAG5-3015, by
NSF grant PHY-9600015,
by a David \& Lucile Packard Foundation Fellowship (to LP), and
by NASA through Hubble Fellowship HF-01084.01-96A from STScI,
operated by AURA, Inc.~under NASA contract NAS5-26555.

\bigskip

\centerline{Table~1 $-$ Correlations with the 19~GHz map}
\smallskip

\begin{center}
\begin{tabular}{lrrrrr}
\hline
\hline
\multicolumn{1}{l}{Template$^{(a)}$}  &
\multicolumn{1}{c}{$\alphaHat^{(b)}$} & 
\multicolumn{1}{c}{$\da$}    & 
\multicolumn{1}{c}{$\damc$}  & 
\multicolumn{1}{c}{$\dar$}   & 
  \multicolumn{1}{c}{$\dT^{(c)}$}    \\
  & & & & & \multicolumn{1}{c}{(\mmicrok)}  \\
\hline
 100\microm  &38.5  &3.3  & 5.8   & 8.3  &138.6$\pm$29.9 \\
 140\microm  &29.8  &2.5  & 4.0   & 6.3  &146.0$\pm$30.9 \\
 240\microm  &46.2  &3.9  & 6.0   &10.0  &143.2$\pm$31.0 \\
 408~MHz     &13.4  &3.1  & 3.4   & 4.4  & 52.3$\pm$17.2 \\
 1420~MHz    & 0.9  &0.1  & 0.2   & 0.3  & 86.5$\pm$28.8 \\
\hline
 100\microm  &47.1  &9.0  &16.2   &17.3  & 65.9$\pm$24.2 \\
 140\microm  &31.6  &6.3  &10.2   &10.3  & 66.4$\pm$21.6 \\
 240\microm  &38.9  &8.4  &12.8   &12.9  & 66.1$\pm$21.9 \\
 408~MHz     & 7.6  &3.7  & 4.7   & 6.2  & 25.1$\pm$20.5 \\
 1420~MHz    & 0.4  &0.2  & 0.2   & 0.3  & 29.8$\pm$22.3 \\
\hline
\end{tabular}
\end{center}

\noindent{\small
$^{(a)}$ 
	 Correlations for $|b|$$>$20$^{\circ}$ (top) and 
         $|b|$$>$30$^{\circ}$ (bottom). \\ 
$^{(b)}$ $\alphaHat$ has units \microk (\mj)$^{-1}$
	 for the DIRBE templates, 
         \mmicrok/K for the 408 MHz template and 
	 \mmicrok/mK for the 1420 MHz template. \\
$^{(c)}$ $\dT \equiv (\alphaHat\pm\dar) \sigma_{Gal}$.
\label{tab:tabCorr}
}






\begin{references} 

 \rn Bennett, C.L., \etal \ 
  1992, ApJ, 396, L7 


 \rn Boggess, N.W., \etal \
  1992, ApJ, 397, 420 

 \rn Boughn, S.P. \etal \
 1992, ApJ, 391, L49

 \rn Brandt, W.N., \etal \
  1994, ApJ, 424, 1
  
 \rn Cottingham, D.A.\
 1987, Ph.D. thesis, Princeton University. 
  
 \rn de Oliveira-Costa, \etal \
  1997, ApJ, 482, L17 (dOC97)

 \rn Draine, B.T., Lazarian, A. \
 1998, ApJ, 494, L19 (DL98)

 \rn Gaustad, J.E., \etal \
 1996, PASP, 108, 351
 

 \rn Haslam, C.G.T., \etal \
  1981, A\&A, 100, 209 

 \rn Kogut, A., \etal \
  1996a, ApJ, 460, 1 (K96a) 

 \rn Kogut, A., \etal \
  1996b, ApJ, 464, L5 (K96b) 

 \rn Kogut, A. \
  1997, AJ, 114, 1127

 \rn Leitch, E.M., \etal \
  1997, ApJ, 486, L23 (L97)

 \rn McCullough, P.R. \
 1997, AJ, 113, 2186
   

 \rn Reich, P., Reich, W. \ 
  1988, A\&AS, 74, 7 

 \rn Simonetti, J.H., \etal \ 
 1996, ApJ, 458, L1
 
 \rn Tegmark, M., Bunn, E. \ 
  1995, ApJ, 455, 1

 \rn Tegmark, M., Efstathiou, G. \ 
  1996, MNRAS, 281, 1297 

 \rn Weiss, R. \
 1980, ARAA, 18, 489 
 
\end{references}
\end{document}